# Imaging the topological current carrying state and the surface to bulk transformation, in Bi$_2$Se$_3$ single crystal and thin film

Amit Jash[1], Ankit Kumar[1], Sayantan Ghosh[1], A. Bharathi[2], S. S. Banerjee[1*]

[1]Department of Physics, Indian Institute of Technology, Kanpur 208016, Uttar Pradesh, India

[2]UGC-DAE Consortium for Scientific Research, Kalpakkam-603104, India

[*]corresponding author email: satyajit@iitk.ac.in.

Magneto-optics based current imaging technique compares the nature of topological current distribution in a single crystal and thin film of topological insulator material, Bi$_2$Se$_3$. The single crystal, at low temperatures, has uniform topological surface current sheets which are about 3.6 nm thick. With increasing temperature, the current partially diverts into the crystal bulk and concomitantly, the sheet break up into a patchy network of high and low current density regions. The temperature dependence of the high current density areas shows that the surface to bulk transformation in the crystal has features of classical phase transition phenomena. The surface area fraction with topological high current density behaves like an order parameter. This phase transition is driven by disorder. In Bi$_2$Se$_3$ thin film we show the presence of quasi one-dimensional topological edge currents which are suppressed with a weak applied magnetic field. The edge current transforms into a uniform bulk current in the film.

Ideal topological insulators (TI) have a topologically protected bulk gapped state enclosed by a conducting surface state [1,2,3,4,5,6]. The high electrically conducting gapless surface states are topologically protected by time-reversal symmetry (TRS). The surface conducting states exhibit exotic features like spin momentum locking which leaves electrical conduction in these materials unaffected by scattering from disorder. The conducting surface states also exhibit Dirac-like linear energy-momentum dispersion [7,8,9,10], chiral spin texture [5] and Landau level quantization [11]. Akin to Si of the semiconductor world, Bi$_2$Se$_3$ is the prototypical material to study TIs. The TI nature of materials has been revealed through observation of Shubnikov–de Haas (SdH) oscillations and weak anti-localization effects found in high field magneto-resistance measurements [12,13,14,15,16].

Scanning SQUID microscopy [17], atom-chip microscopy [18] and scanning photo voltage measurements [19,20,21] have directly imaged the topological edge currents in undoped Bi$_2$Se$_3$ thin films. While all of these imaging studies have been on thin films, there exists no direct imaging of the nature of topological surface current in single crystals of TI. It is known that non-magnetic disorder, viz., selenium (Se) vacancies in $Bi_2Se_3$ electron dopes the bulk [22,23,24,25]. The Se vacancies in $Bi_2Se_3$ TI material result in enhancing the bulk conductivity with respect to the surface [26,27], unlike ideal TI material where the bulk is insulating. How do they affect the topological nature of the material? Recently, non-contact two-coil mutual inductance based bulk susceptibility measurements in $Bi_2Se_3$ single crystals [28,29] showed a predominance of surface conductivity at low temperature ($T$). However above 70 K, the bulk contribution competes with surface conductivity [28], leading to coupling-decoupling effects associated with the high conducting surface states [29]. It would, therefore, be a worthwhile endeavour to image the conducting surfaces in $Bi_2Se_3$ single crystals at different $T$. Using high sensitivity magneto-optical self-field imaging technique, we image and compare the nature of topological currents flowing in a $Bi_2Se_3$ single crystal and thin film. In the crystal at low $T$, electrical conduction is via topological sheet currents of thickness ~ 3.6 nm. With increasing $T$, the current penetrates into the crystal bulk and simultaneously we observe inhomogeneous patches of high and low current density regions develop. Analysis of the temperature dependence of the area of the high current density patches shows that the surface to bulk transformation in the TI behaves like a classical phase transition. The temperature dependence of the area of high current density patches is akin to that of the Ginzburg Landau order parameter. We believe the transition temperature is related to charge doping disorder present in the $Bi_2Se_3$ crystal. In $Bi_2Se_3$ thin film the topological currents are quasi one dimensional edge currents which are suppressed with a low magnetic field of 100 Oe. In the film, the edge current transforms into a uniform bulk current distribution.

We study a single crystal of $Bi_2Se_3$ prepared by slow cooling of stoichiometric melts of high purity Bismuth (Bi) and Selenium (Se) powders. The crystal has dimensions of 1.9 mm × 0.9 mm × 0.02 mm. Crystals from the same batch have already been investigated earlier using electrical transport [24,25] and two-coil mutual inductance [28,29] techniques. The $Bi_2Se_3$ crystal is mechanically exfoliated to obtain a freshly cleaved flat surface on which Cr (5 nm) /Au (50 nm) electrical contact pads are deposited using DC sputtering technique [inset of Fig. 1(a)]. Electrical contacts made using silver paint give a four-probe resistance ~ 10 mΩ (300 K). Epitaxial thin films of $Bi_2Se_3$ of dimensions 2.1 mm × 2.1 mm × 30 nm were grown on STO (111) substrates by RF sputtering (see SI1 and SI2 in supplementary information [30] for X-ray and electrical characterization). The SdH oscillations observed in our single crystal (see

SI3 [30]) confirms that electrical conduction is predominantly via the topological surface states for $T < 25$ K.

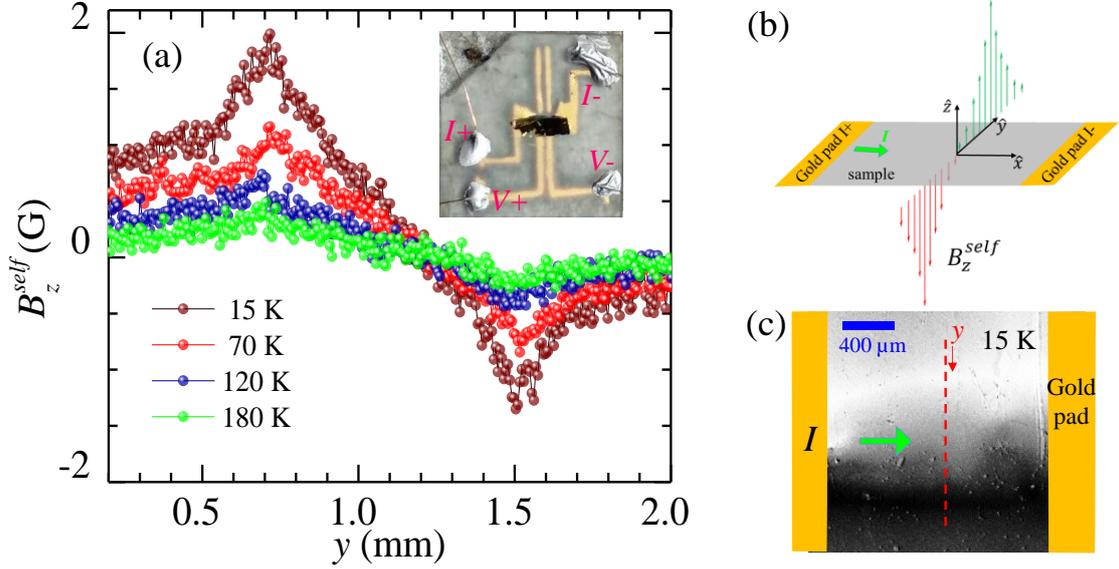

FIG. 1 (a) Figure shows $B_z^{self}(y)$ measured along the red dashed line in Fig. (c) at 15 K, 70 K, 120 K, and 180 K. Inset figure shows the optical image of the $Bi_2Se_3$ single crystal stuck on MgO substrate. (b) Schematic shows the z-component of the self-field distribution, varying along a line across the sample surface with a sheet current flowing along the $\hat{x}$ direction. The length of the red and green vertical arrows schematically represents the variation in negative and positive $B_z^{self}$. (c) Figure shows self-field magneto optical image representing z-component of self-field in the (x,y) plane, $B_z^{self}(x, y)$ measured at 15 K with 35 mA current sent across the $Bi_2Se_3$ single crystal. The location of the current pads is shown by the yellow shaded region. The green arrow represents the direction of the current (I) sent into the sample from the pads. The maximum white and black contrasts are seen near the crystal edges.

We visualize current flow by imaging the spatial distribution of the self-field generated by the current sent across a sample (crystal/film) using the self-field magneto-optical imaging technique (MOI$_{SF}$). The technique has been used [31,32,33] to visualize current distribution in superconductors (for details see Ref. 32). Briefly, MOI$_{SF}$ technique involves high sensitivity spatial mapping of the average Faraday rotation at every pixel in a 512 ×512-pixel view of the sample [using Andor iXon (electron multiplied) EMCCD camera]. The rotation angle is related to the self-field, $B_z^{self}(x, y)$, generated by a current (I) [see schematic in Fig. 1(b)]. Figure. 1(c) is an image of the $B_z^{self}(x, y)$ distribution across the $Bi_2Se_3$ single crystal surface at 15 K with $I = 35$ mA [~ current density (J) of 194 A/cm², assuming a uniform J across the crystal cross-section]. The white and black contrasts correspond to $B_z^{self}(x, y)$ pointing either out of or into the plane of the figure, respectively (supplementary SI4 [30], shows $B_z^{self}$ increases linearly with *I*). Figure 1(a) shows the $B_z^{self}(y)$ profile, measured at different *T* along the red dashed line in Fig. 1(c). At low *T* (15 K) the peaks in $B_z^{self}(y)$ are sharp, however as *T* increases the profile gets rounded off with a decrease in $B_z^{self}$ values.

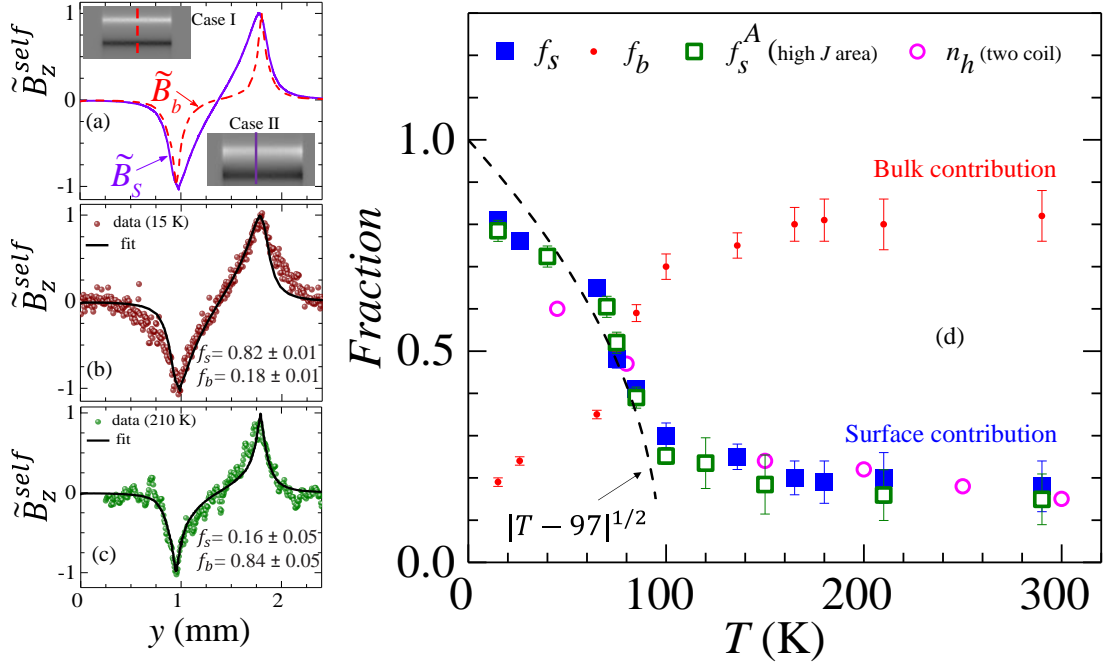

FIG. 2: (a) Figure shows the simulated magnetic field profiles $\tilde{B}_z^{self}(y)$ for bulk (case I) and surface (case II) currents distributed in the crystal. Upper and lower panels are shown as grey scale images, the simulated magnetic field distribution across the crystal, for the case I and II. The behaviour of $\tilde{B}_z^{self}(y)$ and the fitted curve using Eq. 1 (see text) at (b) 15 K, and (c) 210 K. (d) Solid symbols show the variation of calculated surface and bulk fraction with temperature using Eq. 1 (see text). The green square symbol shows the area fraction of the high current density region ($f_s^A$) as a function of temperature (see text for details). Circle symbols show the fraction of high conducting surface state ($n_h$) as a function of temperature determined from the two-coil mutual inductance technique (reproduced from Fig. 5(a) of Ref. [29]). The black dashed line is the fit to the form $f_s^A = 0.101|T - 97|^{1/2}$ (see text for details).

To understand the changes in the $B_z^{self}(y)$ profiles in Fig. 1(a), using COMSOL Multiphysics software we simulate field profile for 35 mA current sent through a 1.6 mm × 0.9 mm × 0.02 mm conductor (same dimension as our crystal) for two cases. Case I : $\tilde{B}_z^{self}(y)$ produced by a uniform current distribution across the sample cross - section (194 A/cm$^2$) and Case II: $\tilde{B}_z^{self}(y)$ produced by two two-dimensional surface current sheets (0.195 A/cm) on crystals top and bottom surface. The calculation was performed using finite current element analysis and the net $\tilde{B}_z^{self}$ was calculated using Biot-Savart's law. The simulated $B_z^{self}(x,y)$ distribution due to bulk current (case I) and surface sheet current (case II) are shown in greyscale in the upper and lower panel of Fig. 2(a), respectively. Figure 2(a) compares the $\tilde{B}_z^{self}(y)$ profile for two cases, using normalized $\tilde{B}_z^{self}(y) = B_z^{self}(y)/\max\{B_z^{self}\}$. The calculated $\tilde{B}_z^{self}(y)$ profiles due to surface current sheet (case II) and bulk current (case I) are represented as $\tilde{B}_s(y)$ and $\tilde{B}_b(y)$ respectively (see Fig. 2(a)). The $B_z^{self}(y)$ profiles at 15 K and 210 K are replotted as $\tilde{B}_z^{self}(y)$ (cf. Figs. 2(b) and 2(c), respectively) and is fitted to Eq. (1),

$$\tilde{B}_z^{self}(y,T) = f_s(T)\tilde{B}_s(y) + f_b(T)\tilde{B}_b(y) \qquad (1)$$

where $f_s(T)$ and $f_b(T)$ are the fraction of current flowing through the samples surface and bulk respectively, with a constraint $f_s(T) + f_b(T) = 1$. Figure 2(b) shows the $\tilde{B}_z^{self}(y)$ profile at 15 K best fits Eq. (1) [black solid line] with $f_s(T) = 0.82$ and $f_b(T) = 0.18$. Thus, at low $T$, the higher $f_s$ value suggests a dominant sheet current flow along the crystal surface with a weaker fraction ($f_b$) of current flowing through the bulk. Compared to 15 K, the fit to Fig. 2(c) at 210 K gives $f_s(T) = 0.16$, $f_b(T) = 0.84$, viz., at high $T$ the fraction of current flowing through the crystal bulk significantly increases compared to that at lower $T$, and the fraction of surface sheet current flow is also lower. Figure 2(d) shows $f_s(T)$ decreasing significantly beyond 80 K, suggesting, there is a transformation from surface to bulk dominated transport. As $B_z^{self}(y)$ is measured near crystal surface, hence the decrease in $B_z^{self}(y)$ with increasing $T$ [Fig. 1(a)] is due to currents penetrating deeper into crystal thereby reducing the self-field strength at the surface generated by these currents. Also note that at 210 K a 16% contribution to surface sheet current flow is still retained, suggesting surface conductivity weakly survives at higher temperature. Note, Fig. 2(d) shows a close match between the values of high conducting surface fraction [$n_h(T)$] determined from two coil ac-susceptibility measurement [29] with our $f_s(T)$.

The measured $B_z^{self}(y)$ is converted into a current density map $J(x,y)$ using a numerical inversion scheme [34]. The resultant $J(x,y)$ map is calibrated using the known current density in the contact pads. The $J(x,y)$ distribution over the crystal surface at 15 K, 100 K, 210 K, and 290 K are shown in Figs. 3(a) to 3(d) in greyscale and colour maps, respectively. Figure 3(a) shows a nearly uniform high $J$ (~ 900 A/cm$^2$) distributed all across the crystal surface at 15 K (note $J$ is lower (~ 200 A/cm$^2$) near the edges as the current sinks deeper into the bulk, possibly due to the presence of crystalline imperfections at the edges). From $T \geq 100$ K [Fig. 3(b) onwards] we see the significant inhomogeneity develop in $J$ distribution which coincides with the increased distribution of current into the crystal bulk [see $f_b(T)$ in Fig. 2(d)]. For $T \geq 100$ K we see that the greenish (and light blue) regions with lower $J$ invade the high $J$ (dark blue) regions of the sample, resulting in an inhomogeneous $J(x,y)$ map [Figs. 3(b)-3(d)]. Figure 2(d) shows $f_s^A$ (= [area of high $J$ (dark blue) regions]/[top surface area]) has identical $T$ dependence as the behaviour of the topological high conducting surface states $f_s$. Thus at high $T$, one has a unique TI state, consisting of topological high conducting surface patches embedded in a low conducting matrix. It appears that with increasing $T$ the patches with topological surface states shrink at the expense of an expanding low conducting phase. Thus at room $T$, while other conventional measurements are unable to detect the fraction of the topological surface state contributing to conductivity or the nature of their spatial distribution in the TI's. We detect about 10% (= $f_s^A$) of topological high surface conductivity patches surviving in Bi$_2$Se$_3$ at room $T$ (Fig. 2(d)). The temperature dependence of $f_s^A$ will be discussed later.

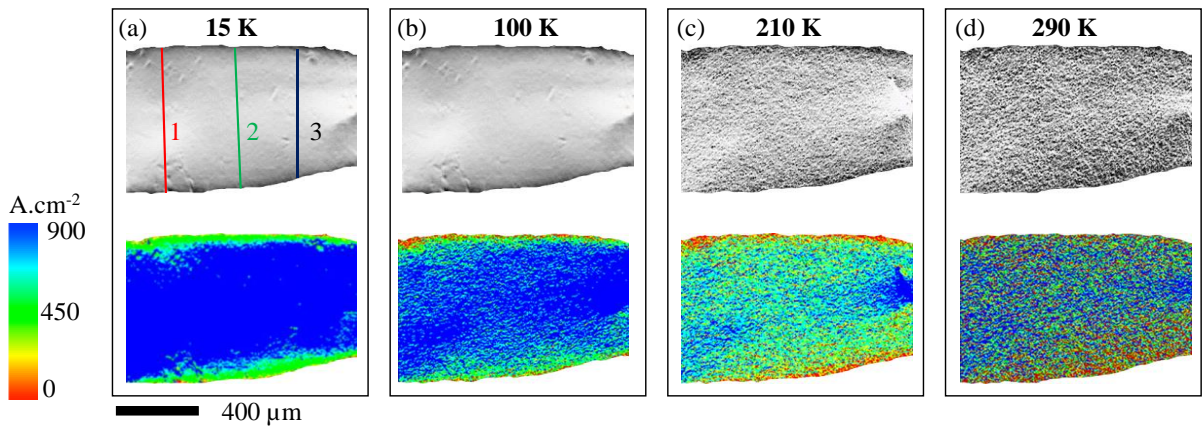

FIG. 3. Shows the $J(x,y)$ distributions in grayscale and RGB scale inside the sample at (a) 15 K, (b) 100 K, (c) 210 K and (d) 290 K (see text for details).

From the $J(x,y)$ distribution at 15 K [Fig. 3(a)] we estimate the average effective thickness ($d_{eff}$) of the topological conducting surface current sheet. A quantity $K$ is determined by integrating $J$ along each of the solid lines marked 1, 2, and, 3 in Fig. 3(a), viz., $K_{1,2,3} = \int_{1,2,3} J \cdot dl$, where $dl$ is the length element along the line. Using, $K \cdot (2d_{eff}) \sim I$ for each of the lines (1,2,3), where $I$ = 35 mA (factor two is for currents distributed along the top and bottom surface sheets in the TI) we get an average $d_{eff} \sim 3.6 \pm 1.0$ nm. This value we have determined for the single crystal is close to the approximately 3 nm thickness of the topological surface state found in thin films of TI [35,36,37].

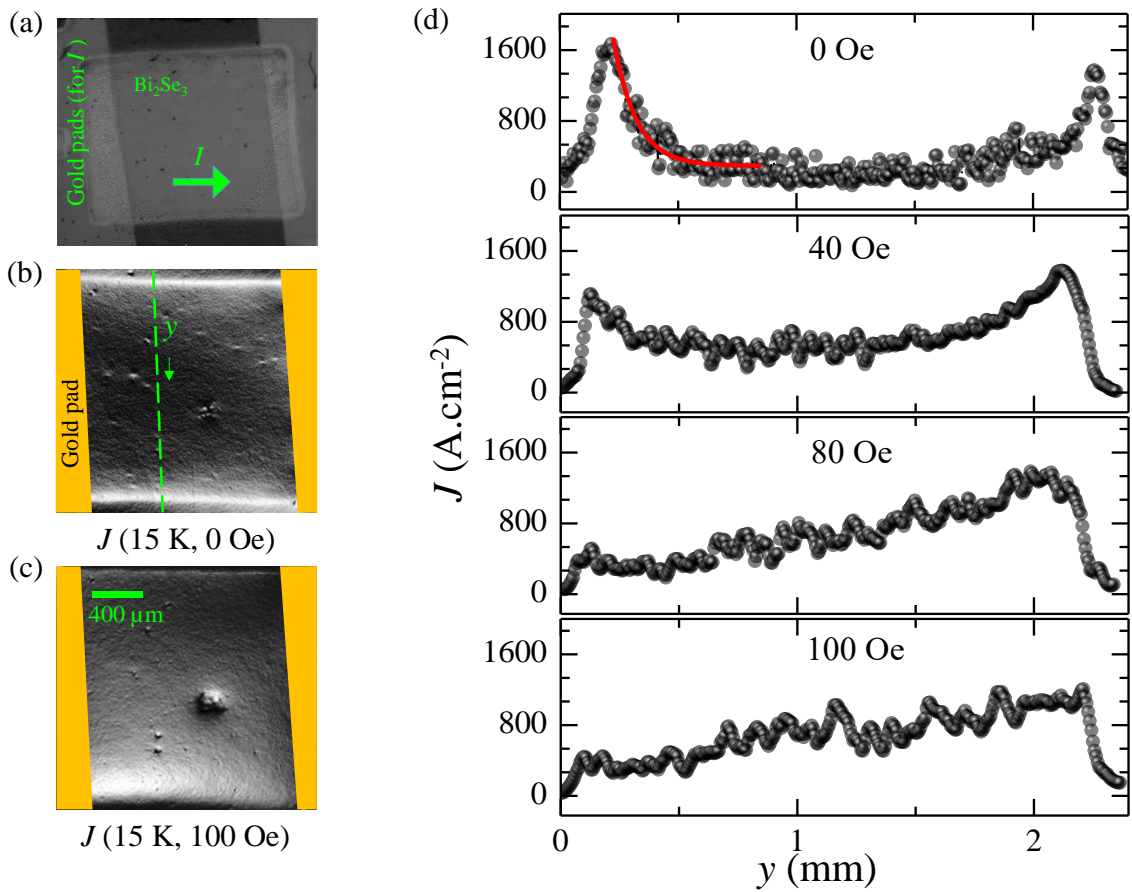

FIG. 4. (a) Figure shows an optical image of Bi$_2$Se$_3$ film with Cr/Au pads. (b) and (c) show $J(x,y)$ images at 15 K for 0 Oe and 100 Oe applied magnetic field respectively. (d) The panel shows the $J(y)$ profiles (measured along the dashed green line in Fig. (b)) at 15 K for 0 Oe, 40 Oe, 80 Oe and 100 Oe. The red solid line is a fit to the 0 Oe data using $J = J_0 \exp(-\frac{y-y_0}{\lambda}) + J_b$, where $J_0 = 1.6 \times 10^3$ A/cm$^2$, $J_b = 260$ A/cm$^2$, $y_0 = 0.22$ mm and $\lambda = 96.70$ μm.

Using self-field images due to 35 mA current sent across a 30 nm thick $Bi_2Se_3$ film, the $J(x,y)$ image in Fig. 4(b) shows a bright contrast indicating strong currents at the film edges. The quasi 1D nature of topological edge current flow $J(y)$ profile in Fig. 4(b) is seen from the strongly peaked nature of $J(y)$ at the film edges which decays exponentially in the film bulk [see the red curve in Fig. 4(d), 0 Oe]. A similar feature of topological currents in thin films of TI has been seen earlier [38,39]. We however additionally show that, with increasing applied field ($H \perp$ to film surface), close to 100 Oe the strong edge currents (at 0 Oe) are suppressed and the current distributes almost uniformly across the film (100 Oe) [see Fig. 4(d)]. As the average current in the bulk increases, a circular defect in the film at 100 Oe becomes visible [Fig. 4(c)] which was indistinct at 0 Oe. With $H$, the degeneracy of the protected edge states ($\pm k$, momentum states) at the Dirac point is lifted, thereby suppressing the gapless edge states in the TI. At 15 K we image $H$ as small as 100 Oe suppresses the topological edge states and redistributes current across the TI. We confirm this through magneto resistance data, which shows a suppression of topological character of the film at 15K with applied $H$ [valid at very low $H$ also, see supplementary SI2 [30]].

It is clear that current flow in the TI thin film is via quasi (1D) one dimensional channels, while in the thick TI single crystals it's a via quasi 2-dimensional sheet current flow. In the quasi 2D TI state of the single crystal, we observe the development of unusual patchy nature of high $J$ state with increasing $T$ beyond 70 K. The observations in fig.3 [as well as fig.2(d)] showing a transformation from a state with a single large area region with high $J$ at low $T$ into patches of low and high $J$ regions, appears like a phase transition in the TI crystal. Figure 2(d) shows that $f_s^A(T)$ behaviour is akin to that of typical Ginzburg Landau order parameter, viz., for $T \to T_c$ = 97±0.5 K (with $T < T_c$), $f_s^A(T) = 0.101|T - 97|^{1/2}$ (dashed curve in fig.2(d)). Non-contact measurements in this $Bi_2Se_3$ crystal [28] have shown the presence of thermally activated delocalization of the charge which have been doped bulk by Se vacancies. The activation of these charges leads to the bulk of the TI crystal to exhibit significant electrical conductivity to appear above ~ 70 K. From a study of the activated nature of electrical conductivity in $Bi_2Se_3$ crystals [28], the activation energy for these doped charges is estimated to typically range between ~ 70 to 100 K. The above $T_c$ is related to this activation energy scale for the charged doped in the TI material bulk by the Se vacancies. Thus, in the novel topological phase of $Bi_2Se_3$ single crystal, we see the emergence of a classical phase transition from a low-temperature phase dominated by large area regions with topological surface state electrons into a higher $T$ phase which is dominated with bulk electrons. The classical phase transition we believe emerges as a result of the disorder, viz., through electron doping Se vacancies present in the TI crystal [16,22,23,28].

In summary, in a thick single crystal of $Bi_2Se_3$ at low $T$ we see quasi 2D uniform current sheets with high $J$ (3.6 nm thick). With increasing $T$ the flow becomes inhomogeneous with patches of high and low $J$. This transformation is concomitant with enhancement in the bulk conductivity due to thermally activated delocalization of Se vacancy doped charges in the bulk. The surface to bulk transformation in the crystal has features of a classical phase transition. In $Bi_2Se_3$ thin film the topologically protected current state is quasi one dimensional which directly transforms into a uniform bulk conducting state. Some of these differences could be related to the difference in the dimensionality of the TI surface state and need future exploration.

SSB acknowledges funding support from DST (AMT-TSDP and Imprint-II programs), IIT Kanpur and help of T. R. Devidas from IGCAR Kalpakam, India and and Hebrew University Jerusalem, Israel. SG acknowledges CSIR, INDIA for funding supports.

**Imaging the topological current carrying state and the surface to bulk transformation, in Bi$_2$Se$_3$ single crystal and thin film**

Amit Jash[1], Ankit Kumar[1], Sayantan Ghosh[1], A. Bharathi[2], S. S. Banerjee[1*]

[1]Department of Physics, Indian Institute of Technology, Kanpur 208016, Uttar Pradesh, India

[2]UGC-DAE Consortium for Scientific Research, Kalpakkam-603104, India

---

[*] Email: satyajit@iitk.ac.in.

## SI1: Bi$_2$Se$_3$ thin film:

The Bi$_2$Se$_3$ thin film has been grown on SrTiO$_3$(111) substrate by the rf magnetron sputtering. A commercially available stoichiometric Bi$_2$Se$_3$ target from ALB Materials Inc of high purity of 99.999% was used as the sputtering target. The distance between the substrate and the sputtering gun was 6.4 cm. The base pressure of the sputtering chamber was kept at $5 \times 10^6$ mBar, and $8 \times 10^3$ mBar argon gas (99.99% purity) pressure was maintained during deposition. The SrTiO$_3$ substrate temperature was kept at 390 °C before and during the deposition [1]. Before deposition, the substrates were carefully cleaned with acetone, alcohol, de-ionized water and purged with nitrogen gas. The rf power was fixed at 30 Watt and average growth rate of Bi$_2$Se$_3$ thin film was 1 nm/min, with film thickness between 5 nm to 40 nm. After deposition, the film was post-annealed in the sputtering chamber in argon environment at temperature 350 °C for four hours. Figure below shows the XRD (X-ray diffraction) pattern only displaying (0 0 3n) direction Bragg peaks of the Bi$_2$Se$_3$ thin film due to the three-fold symmetry of the Bi$_2$Se$_3$ crystal structure, the same as reported in past [1,2]. The XRD direction was carried out in gazing angle mode which avoids the Bragg peak from the SrTiO$_3$ substrate. The direction peaks of (0 0 3n) indicates the rhombohedral structure and the thin film growth along the (111) direction. The sharp XRD peaks indicate the high quality, epitaxial nature of our films.

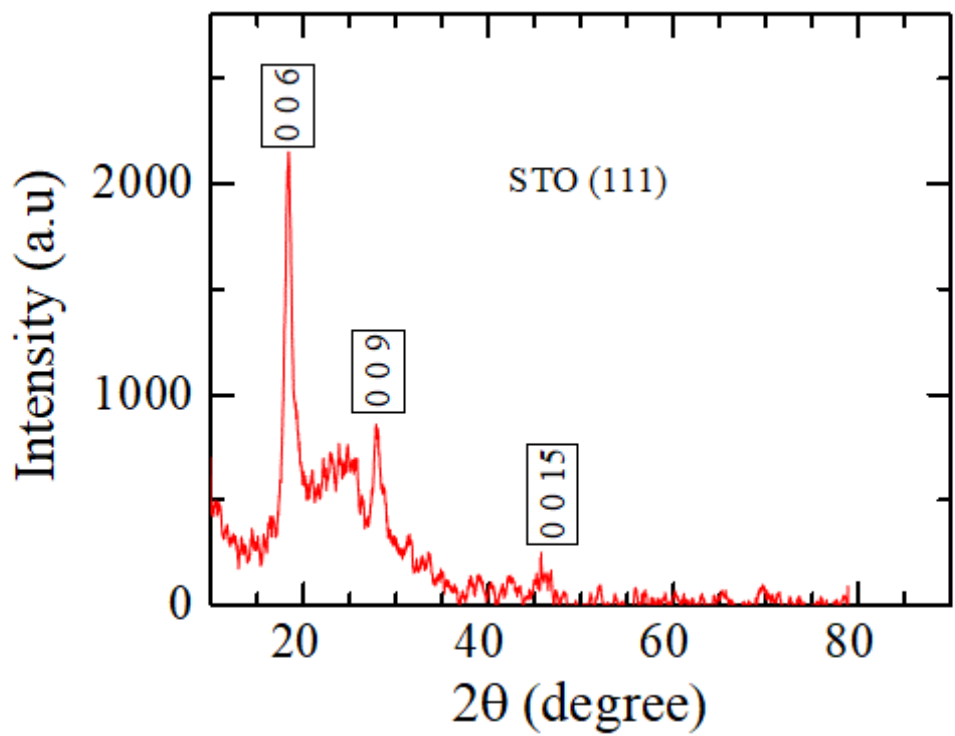

## SI2: Electrical transport measurement of $Bi_2Se_3$ thin film:

Transport measurement of the thin film (2.2 mm × 2.2 mm × 30 nm) was carried out in an APD cryostat in a temperature range of 10 K to 300 K and measurements were done using standard four probe geometry. The figure below shows the bulk resistivity ($\rho$) as a function of temperature ($T$). The resistivity increases monotonically with temperature. This metallic behaviour (of almost linear $\rho$ versus $T$) in $Bi_2Se_3$ over a wide temperature range suggests strong electron-phonon scattering due to the presence of high carrier concentration in the material [3]. Such metallic like conductivity seen in $Bi_2Se_3$ is attributed to the presence of intrinsic defects due to selenium vacancies generated during the growth of $Bi_2Se_3$ thin films [4,5]. These selenium vacancies electron dope the film thereby increasing the carrier concentration in the film.

To better understand this feature, the inset shows the bulk resistivity as a function of film thickness ($d$, for all measured samples) at low temperature (15 K). We know from Fig. 4 of the main manuscript that while significant amount of current flows through the film edges, a significant amount of current also flows through the bulk of the film. Hence the measured $\rho$ will have contribution from the film bulk. The $\rho$ is seen to be strongly dependent on thickness ($d$), viz., $\rho$ decreases with increasing $d$. Note that the resistivity of 5 nm thick film is 4 times larger than the resistivity of 40 nm thick film. The observed thickness dependence of $\rho$ is related to Se vacancies in the film. As the number of Se vacancies in the film increases with $d$, hence the resulting doped charge concentration and the bulk conductivity increases with increasing $d$ (or $\rho$ decreases with increasing $d$). Our result indicates the presence of Se vacancies in $Bi_2Se_3$ thin films.

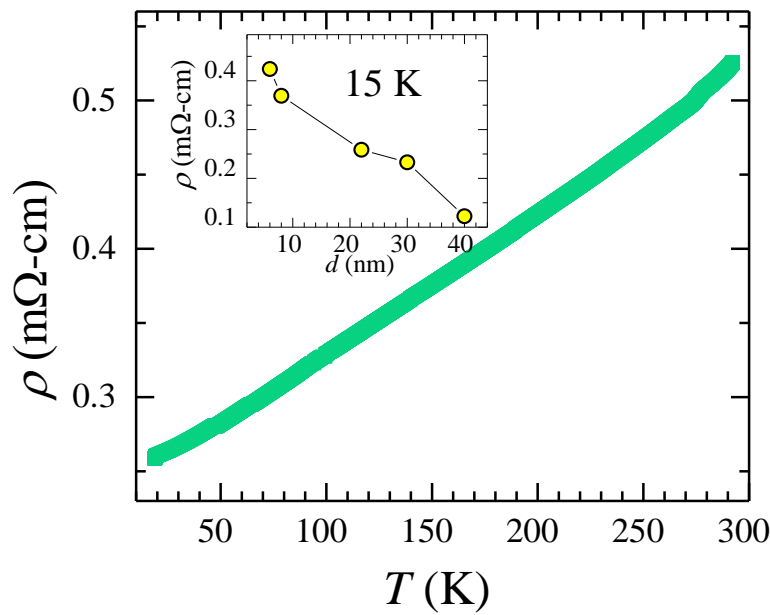

**Magneto conductance behaviour at 15 K in our Bi$_2$Se$_3$ thin film**

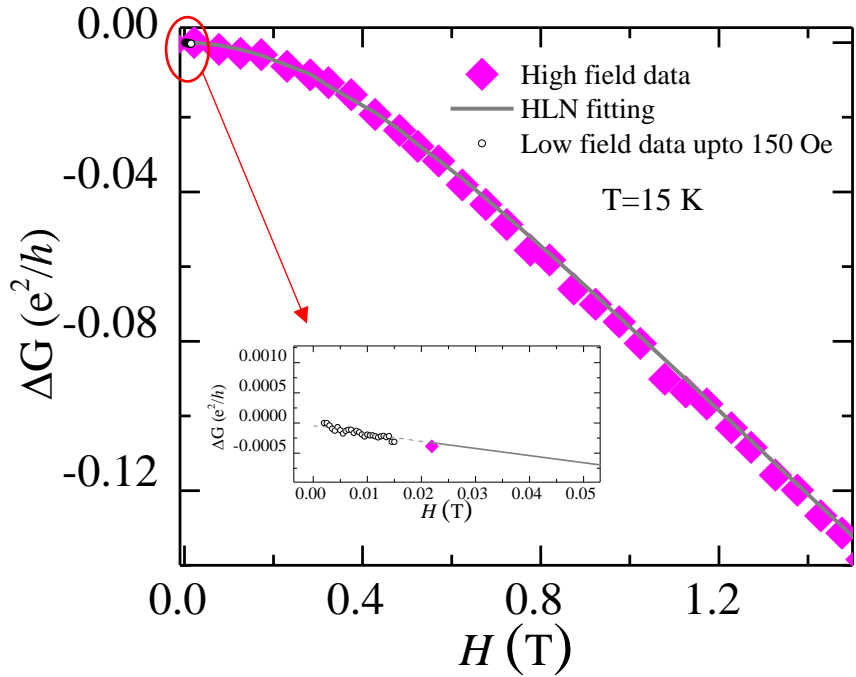

Figure above shows magneto-conductance ($\Delta G$) at 15 K, where $\Delta G = G(H) - G(0)$, which shows the conductance decreases sharply with increasing magnetic field. At the special Dirac point in the Dirac spectrum of the surface states in TI, the momentum states ($+k$ and $-k$) are doubly degenerate. Application of a magnetic field lifts this degeneracy and time reversal symmetry is broken. The spin-momentum locking and presence of time reversal symmetry (TRS), surface state of TI accrues a $\pi$ Berry phase which decreases with the applied magnetic field. Thus, magnetic field supresses the topological gapless conducting edge/surface states. Hence conductance decreases with increasing magnetic field. In magneto-resistance (MR) measurement this shows up as the cusp shaped $\Delta G(H)$ and represents weak anti-localization (WAL) feature, which is a typical feature of a topological insulator. With the application of a field additional channels of electrical transport may open up, which is also deduced from MR measurements.

The magneto resistance of our film was measured using four probe geometry method in Bi$_2$Se$_3$ thin film. High field measurement (solid square symbols, above 220 Oe and upto 6 T) was carried out in PPMS (Quantum Design). For low field measurement below 150 Oe (black circle symbol) measurement was done using an electromagnet (in order to avoid remnant field issues of the superconducting magnet in a PPMS. The remanent field of the superconducting magnet in PPMS when cycled down to 2 Tesla field was found to be 150 Oe. This prevents doing very low field measurements below 150 Oe in a PPMS magnet).

We show the low field ($H$) measurement in the inset. We have fitted the $\Delta G(H)$ data with the well-known Hikami, Larkin and Nagaoka (HLN) model [6,7]:

$$\Delta G = -\alpha \frac{e^2}{2\pi^2 \hbar} \left[ \ln\left(\frac{B_\Phi}{H}\right) - \psi\left(\frac{1}{2} + \frac{B_\Phi}{H}\right) \right],$$ where $\psi$ is digamma function, $e$ is the charge of the electron, $\alpha$ and $B_\Phi$ are the fitting parameters. The fitting is shown by grey solid line with $\alpha$ = 0.89 and $L_\Phi$ = 20 nm. The fitted line for data above 200 Oe is seen to smoothly extrapolate through the $\Delta G(H)$ data taken for $H$ < 150 Oe (see inset). The phase coherence length ($L_\Phi$) is calculated from $B_\Phi$ phase coherence field, viz., $L_\Phi = \frac{\hbar}{4e^2 B_\Phi}$. The value of $\alpha$ gives the effective number of coherent conduction channels. It has been proposed that $\alpha$ = 1/2 for each conduction channel. We have observed at low temperature $\alpha$ is close to 0.5, signifies WAL originates predominantly from the surface Dirac electron [6,7]. At 15 K, $\alpha$ increases from 0.5 to 1 suggesting additional channels of conduction other than topological surface states are active. The $\alpha$ and $L_\Phi$ for different temperature are shown in the table below. Also, the $L_\Phi$ decreases monotonically with increasing temperature due to increased electron-phonon interaction in the bulk state. Our data shows a suppression of the topological surface states with magnetic field as imaged in our magneto-optical measurements.

| Temperature (K) | $\alpha$ | $L_\Phi$ (nm) |
|---|---|---|
| 4 | 0.42 ± 0.05 | 48 ± 1.5 |
| 6 | 0.46 ± 0.05 | 37 ± 1.5 |
| 10 | 0.78 ± 0.05 | 26 ± 1.5 |
| 15 | 0.89 ± 0.05 | 20 ± 1.5 |

## SI3: Observation of SdH oscillation in the Bi$_2$Se$_3$ single crystal and location of the Fermi level within the bulk gap of the material:

Figure (a) shows distinct SdH oscillation in longitudinal magneto-resistance ($\Delta R_{xx}$ vs magnetic field ($B$)) measurements at 10 K using standard Van der Pauw geometry. This data is reproduced from Ref. [8] (Ref. 28 in main MS). $\Delta R_{xx}$ is calculated by subtracting from the experiment $R_{xx}(B)$ values with a polynomial fit to the data ($R_{poly}(B)$), i.e, $\Delta R_{xx} = R_{xx}(B) - R_{poly}(B)$. The polynomial form of $R$ is $R_{poly}(B) = R_0 + R_1 B + R_2 B^2$, where $R_0 = 9.26 \times 10^{-3}$ $\Omega$, $R_1 = -1.71 \times 10^{-6}$ $\Omega \cdot T^{-1}$, $R_2 = 2.91 \times 10^{-5}$ $\Omega \cdot T^{-2}$. Upper inset shows the variation of $R_{xx}$ as a function of magnetic field $B$ measured at 10 K. Lower inset shows the SdH oscillation at different low temperature.

We use Lifshitz-Kosevich (LK) equation: $\Delta R_{xx} = \sqrt{0.011B} \left( \frac{\frac{11.12}{B}}{\sinh\left(\frac{11.12}{B}\right)} \right) e^{-\frac{19.38}{B}} (0.95) \cos\left[ 2\pi \left( \frac{F}{B} + \beta \right) \right]$ to analyse the SdH oscillation seen in our Bi$_2$Se$_3$ sample, where $F$ and $\beta$ are the fitting parameter. From the oscillation period ($F$) of the SdH oscillations seen in the transport data, the measured surface carrier density per area is found to be $n_s = (2.268 \pm 0.012) \times 10^{12}$ cm$^{-2}$, which corresponds to a Fermi wavevector for the 2D surface state to be $k_F = \sqrt{2\pi n_s} = (0.0377 \pm 0.0027)$ Å$^{-1}$. By placing the location of $k_F$ on the ARPES spectrum of Bi$_2$Se$_3$ (see $k_F$ marked by a vertical yellow line in the Fig. (b) which is Fig. 1 ARPES data of Bi$_2$Se$_3$ as published by M. Bianchi et al., Nature Commun. **1**, 128 (2010)), we see that the Fermi energy is approximately 30 meV above the Dirac point and about 100 meV below the bottom of the bulk conduction band (shown by green dash line). The position of the Fermi level suggests the carriers are from surface states and not from bulk bands. The charge carriers from bulk conduction do not contribute which suggests we do not have a non-topological two-dimension gas contributing to conductivity. Hence, the sheet current distributions at low temperature arises due to the topological surface current of Bi$_2$Se$_3$ single crystal (Fig. 3(a) in MS).

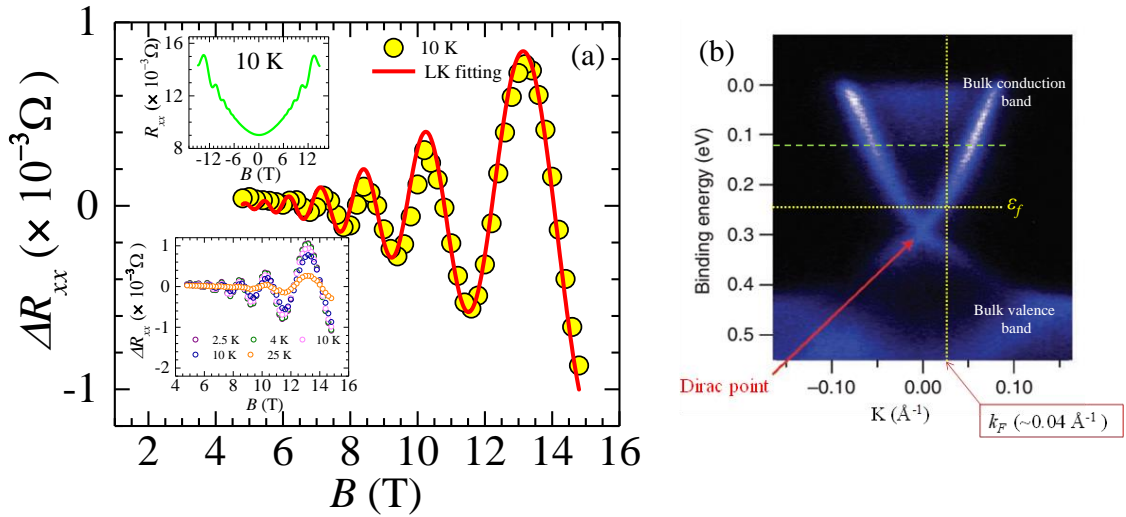

## SI4: Self-field due to different amplitude of current:

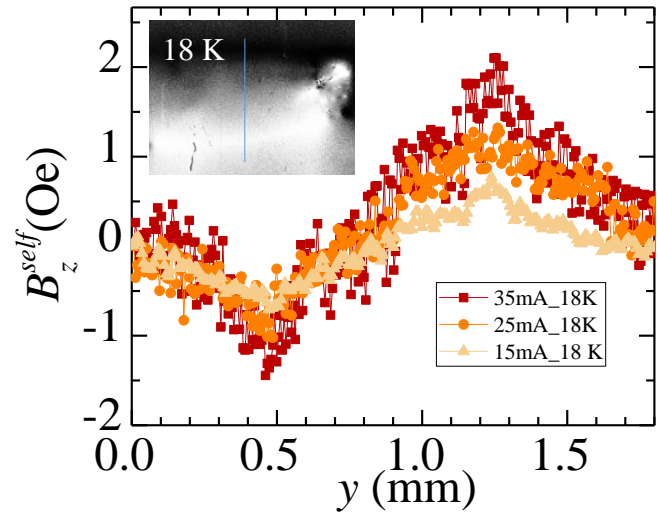

Figure above shows the magnetic field profiles at 18 K for different applied current through the $Bi_2Se_3$ sample. Line scan is taken along the solid line to each $B_z^{self}(x, y)$ image at different current. The $B_z^{self}(x, y)$ at 35 mA is 2.05 Oe (at the edge) which decreases to 0.60 Oe when 15 mA current is applied through the sample. The self-magnetic field strength increases linearly with the applied current which justifies the linearity response of driving current to the captured $B_z^{self}$ distributions.